\documentclass{aa}  
\newcommand{\Msun}{\thinspace\hbox{$\hbox{M}_{\odot}$}}

\newcommand{\Lsun}{\thinspace\hbox{$\hbox{L}_{\odot}$}}

\usepackage{graphicx}
\usepackage{txfonts}
\begin{document}

\title{Optical detection of the radio supernova SN 2000ft in the circumnuclear region of the
          luminous infrared galaxy NGC 7469\thanks{Based on observations with the NASA/ESA Hubble Space
          Telescope at the Space Telescope Science Institute, which is operated by the
          Association of Universities for Research in Astronomy, Inc., under NASA
          contract NAS5-26555.}}

	  \titlerunning{L. Colina et al.: Optical detection of SN 2000ft in NGC 7469}
	  
%   \subtitle{I. Overviewing the $\kappa$-mechanism}

   \author{L. Colina\inst{1}, T. D\'{\i}az-Santos\inst{1}, 
              A. Alonso-Herrero\inst{1},
              N. Panagia\inst{2,3,4},
              A. Alberdi\inst{5},  J.M. Torrelles\inst{6},  \and
              A.S. Wilson\inst{7}}

   \offprints{L. Colina}

   \institute{Instituto de Estructura de la Materia (CSIC-IEM), Serrano 121, 28006 Madrid, Spain \\
	     \email{colina, tanio, aalonso@damir.iem.csic.es}
	     \and
          Space Telescope Science Institute, 3700 San Martin Drive, Baltimore, MD 21218, USA 
	    \and
          Istituto Nazionale di Astrofisica (INAF), Via del Parco Mellini 84, I-00136, Rome, Italy
	   \and
         Supernova Ltd., OYV \#131, Northsound Road, Virgin Gorda, British Virgin Islands\\
            \email{panagia@stsci.edu}
          \and
          Instituto de Astrof\'{\i}sica de Andalucia, CSIC-IAA, Apartado 3004, 18080 Granada, Spain\\
	    \email{antxon@iaa.es}
	   \and
           Instituto de Ciencias del Espacio (CSIC)-IEEC, Facultat de F\'{\i}sica, Planta 7a, Universitat de Barcelona, 
	    Av. Diagonal 647, 08028 Barcelona, Spain\\
	    \email{torrelles@ieec.fcr.es}
	    \and
           Department of Astronomy, University of Maryland, College Park, MD 20742, USA \\
	    \email{wilson@astro.umd.edu}
             }

   \date{Received December 29, 2006; accepted March 12, 2007}

% \abstract{}{}{}{}{} 
% 5 {} token are mandatory
 
  \abstract
   {(Ultra)Luminous Infrared Galaxies (ULIRGs) produce massive stars in large quantities in their central 
   starburst regions. The expected rate of supernova explosions is, on average, about one per year. 
   Detection of these supernovae in the expected numbers has proven elusive so far.}
    {Illustrate the general benefits and limitations of supernova searches in (U)LIRGs in the optical 
    using the previously detected luminous type II radio supernova SN 2000ft in NGC 7469 as a study case.}
    {Multi-epoch Hubble Space Telescope (HST) optical imaging at different wavelengths and 
    two dimensional PSF fitting algorithms are used to search, and characterize, the optical emission of SN 2000ft.}
   {SN 2000ft is detected in two independent Planetary Camera images (F547W and F814W)  
    taken May 13, 2000,  about two months before the
     predicted date of the explosion (July 19, 2000),  based on the
     analysis of its radio light evolution by Alberdi and collaborators.
    The apparent optical magnitudes and red color of SN 2000ft indicate that it
     is observed through an extinction of at least A$_V$= 3.0 magnitudes.
     The extinction corrected lower limit to the absolute
     visual magnitude (M$_V$ $\leq -$ 18.0),  identifies SN 2000ft as
     a luminous supernova in the optical, as other luminous
     radio supernovae before.  SN 2000ft exploded in a region located
      at only 0\farcs1 (i.e. 34 $\pm$ 3 pc) west of a faint cluster (C24). No parent cluster is
      identified within the detection limits of the HST short exposures.}
   {The unambiguous detection of SN 2000ft in
      the visual shows that multi-epoch sub-arcsecond (FWHM $\sim$ 0\farcs1) optical imaging is also a valid tool
      that should be explored further 
     to detect supernovae in the dusty (circum)nuclear regions of (U)LIRGs.}


\maketitle
%
%________________________________________________________________

\section{Introduction}

A large fraction of the star formation in the Universe has taken place in optically-obscured
galaxies (Lagache, Puget \& Dole 2005). At low-$z$, luminous infrared galaxies (LIRGs) with luminosities
L$_{IR}(8-1000~\mu$m)$>$10$^{11}\Lsun$, are more numerous than optically-selected
starburst and Seyfert galaxies of comparable bolometric luminosity, and at the highest
luminosities, L$_{IR}>10^{12}\Lsun$, they exceed the space density of
quasi-stellar objects (QSOs) by a factor of 1.5--2 (Sanders \& Mirabel 1996
and references therein). In a large fraction of low-{\it z} LIRGs the bulk of the star formation is
concentrated in (circum)nuclear ring-like or mini-spiral structures (Alonso-Herrero et al.
2006 and references therein) characterized by the presence of numerous dusty, young
($\leq$ 100 Myr), massive ($\sim$ 10$^{5}-10^7$ \Msun) star clusters and HII regions (Scoville et al. 2000, Alonso-Herrero et al.
2002, 2006).

Searches for supernovae associated with recent star formation processes in LIRGs have proceeded with various rates of success. Multi-epoch Very Long Baseline
 Interferometry (VLBI) radio imaging (Lonsdale et al. 2006, Smith et al. 1998) has
  provided evidence for numerous
  radio supernovae in Arp 220. Candidates to radio supernovae have also been detected in LIRGs such as NGC 6240
   (Gallimore \& Beswick 2004) and Arp 299 (Neff et al.
2004) based on Very large Array (VLA) and VLBA imaging.  Attempts to detect the expected high rate of
supernovae in LIRGs have also been conducted in the near-infrared from the ground
(Maiolino et al. 2002, Mannucci et al. 2003, Mattila et al. 2004, Mattila \& Meikle 2001) and with NICMOS on the
HST (Cresci et al. 2006).
The observed near-IR
supernova rate is a factor 3 to 10 smaller than estimated from the far-infrared
luminosity (Mannucci et al. 2003). Thus the conclusion that dust extinction in these galaxies is so high that obscures most SN even in the near-IR seems unavoidable.
However, the distribution of dust in the central regions of (U)LIRGs is clumpy on scales of 100 pc or less (Alonso-Herrero et al. 2006) and consequently star-forming regions with relatively low extinction (A$_V \leq$ 2-6 mag) coexist with highly absorbed young star clusters (Scoville et al. 2000; Surace, Sanders \& E
vans 2000). Therefore, the possibility of detecting supernovae in the low-extinction regions is still open and should be explored further
with high angular resolution imaging.

NGC 7469 is a LIRG with
L$_{IR}$ = 5 $\times$ $10^{11} \Lsun$, at
a distance of 70 Mpc  (Alberdi et al. 2006). NGC 7469 contains a luminous
Seyfert 1 nucleus surrounded by a dusty circumnuclear starburst ring of about 1.6 kpc in
diameter (Cutri et al. 1984; Wilson et al. 1986; Wilson et al. 1991; Miles, Houck \&
Hayward 1994; Genzel et al. 1995; Malkan, Varoujan \& Tam 1998; Scoville et al. 2000).
SN 2000ft was detected
at 8.4 GHz on October 27, 2000 (Colina et al. 2001a,b) during the course of a monitoring campaign designed to search for radio supernovae in NGC 7469. Detailed
subsequent modelling of the radio multi-frequency light curve has allowed to
establish that SN 2000ft displays the radio properties of a bright type II supernova that exploded on July 19, 2000 (Alberdi et al. 2006, 2007). Previous attempts at detecting SN 2000ft in the optical from the ground failed (Li et al. 2001).
This letter presents the HST detection in the optical of supernova SN 2000ft in NGC 7469. Implications for the detection of supernovae in (U)LIRGs using
multi-epoch sub-arcsec optical imaging are discussed.

\section{Hubble Space Telescope imaging and photometry}

    Optical HST images of NGC 7469 were retrieved from the
     archive and calibrated using the {\it On the Fly Recalibration} (OTFR) system.
     Images were taken with the WFPC2 Planetary Camera (F547M, F606W, and F814W),
     and with the ACS High Resolution Camera (F330W). These images have a plate
     scale of 0\farcs0455 (WFPC2/PC) and 0\farcs026 (ACS/HRC), providing
     a PSF (TinyTim FWHM) of between 0\farcs051 (F330W) and 0\farcs085 (F814W).
      The full set of HST data also includes lower angular
     resolution ultraviolet (WFPC2/F218W) and near-infrared (NICMOS, Scoville et al. 2000) imaging, that
     have been used in the modelling of the properties of the circumnuclear star clusters
     (see D\'{\i}az-Santos et al. 2007 for details).

     Accurate photometric measurements of point-like sources in complex local backgrounds as that present
     in the nuclear regions of NGC 7469 cannot rely on standard aperture photometry.  The
     flux contribution of the background changes from source to source, and therefore
     has to be modelled and subtracted accordingly. The fluxes and magnitudes reported in this
     paper make use of a $\chi^2$-minimizing Marquardt algorithm that models point-like sources
     on top of a diffuse background emission as the combination of a two-dimensional Gaussian and
     planes functions representing the point source and surrounding background, respectively
     (D\'{\i}az-Santos et al. 2007). Filter-dependant PSF corrections were done assuming realistic PSF
     TinyTim models (Krist et al. 1998)

     \section{Results and Discussion}
     
     \subsection{Optical Detection of  Supernova SN 2000ft}

Supernova SN 2000ft was first detected at 8.4 GHz on October 27, 2000 (Colina et al.
2001a,b). Subsequent monitoring at various radio frequencies allowed us to investigate in detail the radio light curve of SN 2000ft and to predict July 19th, 2000
 ($\pm$ 40 days) as the  date of the explosion (Alberdi et al. 2006, 2007). Optical HST images of NGC 7469 were
  taken for different programs at three different epochs (referred hereinafter as epoch 1, 2 and 3) from June 10,
   1994 (F606W) to May 13, 2000 (F547M and F814W), and November 20, 2002 (F330W). In particular, epoch 2
  images were taken about two months before the explosion date predicted from the radio light curve analysis. On
   the other hand, epoch 1 and 3 images were obtained about 6 years before and 2.5 years after the explosion,
    respectively.

Since the angular resolutions of the A configuration VLA and HST images are similar to within a factor of two, it is assumed that the bright radio and optical Seyfert 1 nucleus are spatially coincident. The comparison of the radio map and optical images shows the presence of an optical point source in the epoch 2 images
 (F547M and F814W) in a region of the circumnuclear star-forming ring coincident with the position of SN 2000ft,
  and close to cluster C24 (see Figure 1, and D\'{\i}az-Santos et al. 2007 for the cluster naming convention). This
   point source is not detected in epochs 1 (F606W)  and 3 (F330W) images indicating that the optical emission 
    due to a transient source. Moreover, since the epoch 2 images were taken about two months before the
    predicted epoch of the explosion, the increase in flux must represent the transient optical emission of the radio
     supernova SN 2000ft.

%                                     Two column figure (place early!)
%______________________________________________ Gamma_1 (lg rho, lg e)
   \begin{figure*}
   \centering
    \includegraphics[width=16cm]{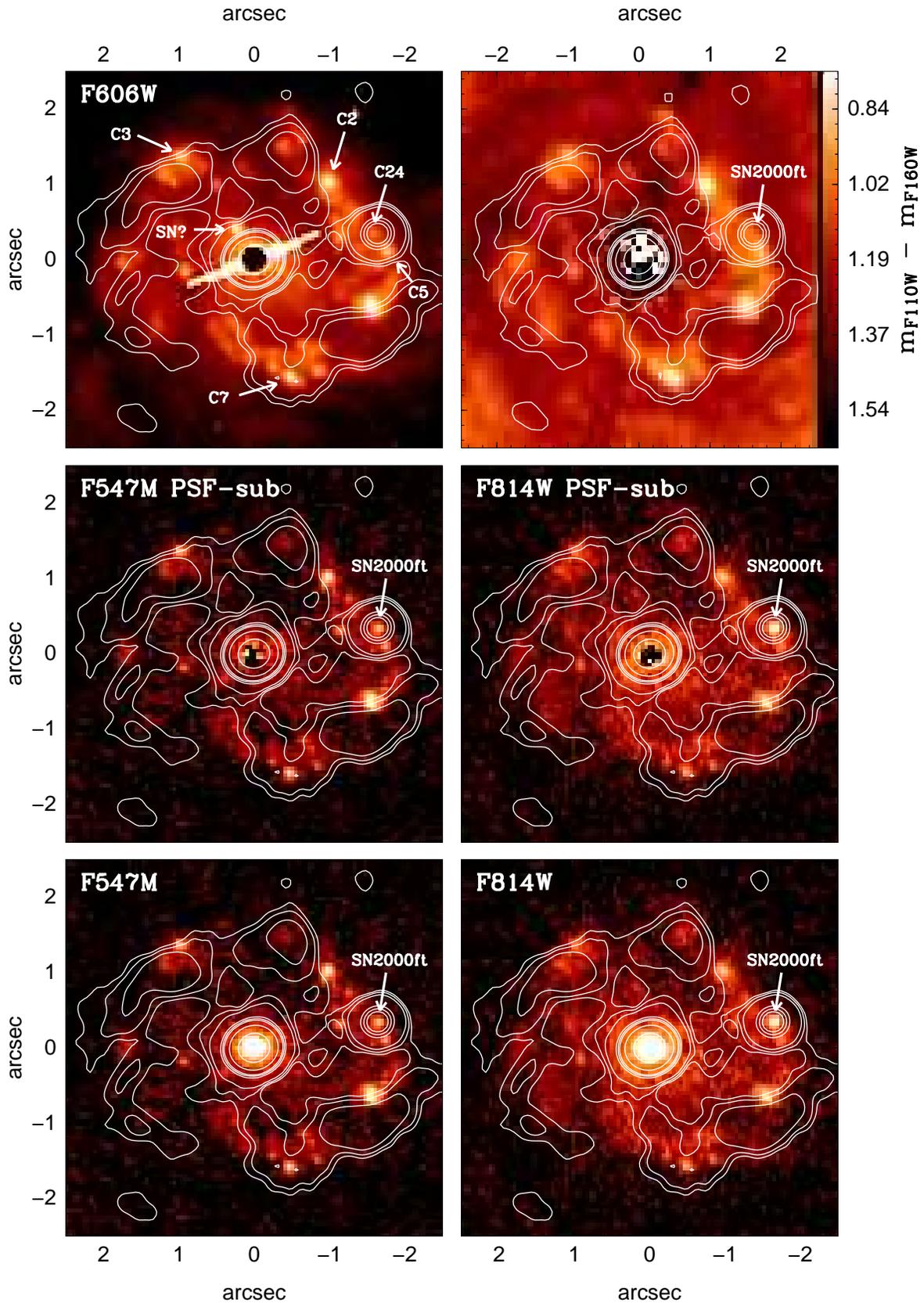}
   \caption{HST WFPC2 Planetary Camera images of the nucleus and circumnuclear star-forming ring of NGC 7469 in
the available optical filters.  Contours represent the VLA 8.4 GHz image taken on October 27, 2000 when the supernova SN 2000ft was first detected (Colina et al. 2001). The HST images taken
 in May 2000 (F547M and F814W) clearly show an increase in flux in the region where SN 2000ft exploded with
  respect to that of clusters C2, C3 and C5 as observed
 through the F606W filter in 1994
(see text and Table 1). Image F606W shows a bright source (SN?) at 0\farcs47 from the nucleus that could be real and associated with another supernova (see section $\S$3.1 for details). The dust distribution is shown in the HST/NICMOS J$-$H color image (F110W $-$ F160W, upper right panel) taken in Nov. 1997. Assuming the diffuse stellar emission is associated with an old population, the range of colors corresponds to visual extinctions of $0 < A_V \leq 3$ mag (between 1 and 5 mag if a young stellar population is considered). The bright Seyfert 1 nucleus has been removed from the images in the top and central panels to better identify the surrounding star clusters and diffuse structure in the ring, in particular in the F606W and J$-$H color images. The original F547M and F814W images with the bright nucleus are presented in the bottom panels for reference.}
              \label{FigGam}%
    \end{figure*}

To establish the reality of the optical transient, the increase in flux associated with SN 2000ft has been measured relative to that of non-variable bright star
clusters like C2, C3 and C5 located in the ring (see Figure 1). Since no pre-explosion images are available in the F547M and F814W filters,  the fluxes of the different clusters and SN 2000ft are
normalized to the epoch 1 F606W flux (see Table 1). 
The differences in the measured F547M and F814W to F606W fluxes of the clusters are due to differences
in the bandpass of the filters (pivot wavelengths of 5483\AA, 8012\AA, and 5997\AA, respectively), and to intrinsic cluster to cluster changes in their spectral energy distribution due to variations in internal extinction and ages. All these effects combined result in F547M and F814W to F606W flux ratios that could be different from one. In fact, the average values (see Table 1) indicates that the F547M and F606W fluxes of the nearby star clusters are about the same (1.1 $\pm$ 0.2) while the F814W flux is about half (0.51 $\pm$ 0.04). However, the F547M and F814W flux associated with with the region where SN 2000ft exploded is about 5 and 8 times that of the F606W flux, respectively (see Table 1 for more accurate values). Therefore, relative to the non-variable bright star clusters, the optical flux associated with SN 2000ft increased the flux in the region where it exploded by factors of about 4 and 15 in the F547M and F814W light, respectively.

We should also note that image F606W taken in June 1994 shows the presence of a bright source at  0\farcs47 northeast of the Seyfert 1 nucleus (object SN? in the F606W image in Figure 1). This source is not detected in the images F547M and F814W taken in May 2000, about six years later.  It is unlikely that the source is an artifact.
Ghosts of  bright sources like the Seyfert 1 nucleus are known to exist. However,
field flattener ghosts are found along a line from the center of the Planetary Camera to the bright object (nucleus),
and away from the center. This is not the case for this object. Filter ghosts can also be ruled out since they are usually fan-shaped and typically about 1\% of the bright source (nucleus) countrate. The source is too bright and also point-like (0\farcs1 FWHM).  The high counts (113000 electrons) also rules out cosmic rays and hot pixels. Cosmic ray events produce a well defined distribution of the number counts per event with no single events producing more than 10000 electrons (Heyer et al. 2004), i.e. a factor ten less than measured.  
Therefore if the object were real, it could have been a supernova that exploded at a distance of only 160 pc northeast of the Seyfert 1 nucleus with an apparent F606W magnitude of 19.1.  No other image of NGC 7469 were taken close enough in time to confirm or disprove the reality of this detection. 

%_____________________________________________________________
%                                             Two column Table 
%_____________________________________________________________
%
\begin{table*}
\begin{minipage}[t]{\columnwidth}
\caption{HST Photometry for SN2000ft and reference star clusters}             
\label{catalog}
%\label{table:1}      
\centering         
\renewcommand{\footnoterule}{} % to avoid a line before footnotes 
\begin{tabular}{c c c c c c c c}     % 8 columns 
\hline\hline       
Region & ${\rm F330W}$ & ${\rm F547M}$ & ${\rm F606W}$ & ${\rm F814W}$ & 
\normalsize{$\frac{\rm F547M}{\rm F606W}$} & \normalsize{$\frac{\rm F814W}{\rm F606W}$} & 
\normalsize{$\frac{\rm F330W}{\rm F606W}$} \\
(1)\footnote{Cluster label after D\'iaz-Santos et al. (2007)} & (2)\footnote{Measured fluxes in columns (2) to (5) given 
in erg~s$^{-1}$~cm$^{-2}$~\AA$^{-1}$} & (3) & (4) & (5) & (6)\footnote{Flux ratios for the different regions
as obtained from the measured fluxes given in previous columns} & (7) & (8) \\
\hline                    
C2 &   1.69$\times$10$^{-16}$ &  1.05$\times$10$^{-16}$ &  8.85$\times$10$^{-17}$ &  4.95$\times$10$^{-17}$ &      1.2 &      0.6 &      1.9 \\

C3 &   7.38$\times$10$^{-17}$ &  6.74$\times$10$^{-17}$ &  4.94$\times$10$^{-17}$ &  2.36$\times$10$^{-17}$ &      1.4 &      0.5 &      1.5 \\

C5 &   6.42$\times$10$^{-17}$ &  4.03$\times$10$^{-17}$ &  4.67$\times$10$^{-17}$ &  2.27$\times$10$^{-17}$ &      0.9 &      0.5&      1.4 \\

Mean & \dots & \dots & \dots & \dots & 1.1$\pm$0.2 & 0.51$\pm$0.04 & 1.6$\pm$0.2\\

\hline

C24\footnote{Photometry of the closest cluster (C24) identified at a distance of 0\farcs1 from SN 2000ft. No measurement is given for filters F547M and F814W due to its
close distance to SN 2000ft.} &   3.54$\times$10$^{-17}$ & \dots &  1.60$\times$10$^{-17}$ &   \dots
&  \dots &   \dots &    2.2 \\

C24(fit)\footnote{Flux values for the corresponding filters, obtained from the best-fitted spectral energy distribution to the cluster C24 photometry covering the ultraviolet to near-infrared range (D\'iaz-Santos et al. 2007).} & 2.90$\times$10$^{-17}$ & 1.97$\times$10$^{-17}$ & 1.63$\times$10$^{-17}$ & 1.01$\times$10$^{-17}$ & 1.2 & 0.6 & 1.8 \\

SN\_Region\footnote{F330W and F606W values represent the fluxes measured
through an aperture of 0\farcs14 diameter centered at the position where SN 2000ft
is detected in the F547M and F814W images.} & 1.48$\times$10$^{-17}$ & \dots & 1.43 $\times$10$^{-17}$ & \dots & \dots & \dots & 1.0 \\

SN 2000ft\footnote{Photometry of SN 2000ft in F547M and F814W images taken in May 13, 2000} & \dots & 6.83$\times$10$^{-17}$ & \dots & 1.08$\times$10$^{-16}$ & 4.8 &
     7.6 & \dots \\
\hline                  
\end{tabular}
\end{minipage}
\end{table*}

\subsection{SN 2000ft as a Bright Optical Type II Supernova}

The increase in flux detected near cluster C24 (Fig.1) in
the epoch 2 images is interpreted as the optical emission from supernova SN 2000ft,
with apparent magnitudes 19.29 $\pm$ 0.17 and 17.55 $\pm$ 0.14 in the F547M and F814W filters, respectively. Taking into account the F547M to F555W magnitude
offset (0.006 mag) and the HST to ground-based V and I magnitude conversions for the
Planetary Camera (Holtzman et al. 1995), the HST magnitudes translate to V and I magnitudes of
m$_V$= 19.20 $\pm$ 0.17 and m$_I$= 17.45 $\pm$ 0.14.

The absolute optical magnitude of SN 2000ft can be derived after reddening corrections based on
the comparison of the observed and expected intrinsic V$-$I colors are applied.  The detailed analysis of the SN 2000ft radio emission (Alberdi et al. 2006), allowed us to confirm that the shape of its radio light curve is not consistent with that of any known SNIb/c (Weiler et al. 2002) whereas it behaves like a typical
type II supernova.  Moreover, its peak radio luminosity identifies SN 2000ft as luminous radio supernova 
(Alberdi et al. 2006).

 The epoch 2 optical images were taken on May 13, 2000, about two months before the predicted explosion date
 (July 19th, 2000 $\pm$ 40 days; Alberdi et al. 2006, 2007). Therefore within the uncertainties of the
 prediction, the optical detection of SN 2000ft appears to be close to the explosion date. Type II supernovae have
 intrinsic optical colors of V$-$I $\leq$ 0.5 during the early 50 days after the explosion (Vink\'o et al. 2006,
Zhang et al. 2006,  Ho et al. 2001, Fassia et al. 2000).Therefore, the intrinsic V$-$I color of SN
2000ft is assumed to be 0.5, or less. Since the measured V$-$I color for SN 2000ft is 1.75 $\pm$ 0.22,  this color excess
 translates into extinctions of at least 3.0$\pm$0.5 and
 1.8$\pm$0.3 mag in the V and I bands, respectively. The corresponding V and I extinction corrected absolute
magnitudes are $\leq -$18.0 ($\pm$0.6) and $\leq -$18.5 ($\pm$0.3), respectively.  

A relatively large fraction (7 out of 24) of type II supernovae classified as luminous have an average {\it B} 
magnitude of $-$18.8 ($\pm$0.6) at their maximum light, while more regular ones are about two magnitudes 
fainter with an average magnitude of $-$16.6 ($\pm$0.6).  
Considering that type II supernovae have an average B$-$V color of 0.0 at maximum light (Patat et al. 1994), the visual ({\it V}) absolute magnitude of the luminous supernovae correspond to $-$18.8. Therefore since SN 2000ft  is brighter than $-$18.0 in the visual, it can be identified as a luminous type II supernova in the optical. Moreover, prototypes of these bright optical type II supernovae include SN 1979C and SN 1988Z (Patat et al. 1994) which also belong, as SN 2000ft, to
the class of luminous type II radio supernovae (Alberdi et al. 2006 and references). In conclusion, SN 2000ft is both a luminous optical and radio type II supernova.

\subsection{SN 2000ft in a Dust-obscured Massive Star Cluster}

To establish the exact location of SN 2000ft, the relative astrometry of the peak of the SN 2000ft optical emission with respect to that of several bright point-
like clusters (clusters C2, C3, C5 and C7 in Figure 1) has been
independently measured in the F547M and F814W images.  In addition, the same measurements have been performed for the cluster C24 using the F606W image.  Positional measurements were done with different methods including TinyTim PSF fitting (D\'{\i}az-Santos et al. 2007), Gaussian and intensity weighted centroid algorithms. These methods give consistent results with positional uncertainties of a tenth of a Planetary Camera pixel  (i.e. $\leq$ 0\farcs00455). This high precision relative astrometry allows us to pin down the position of SN 2000ft at a location of
0\farcs100 $\pm$ 0\farcs009 (34 $\pm$ 3 pc) west of the faint, unresolved (FWHM
$\leq$ 0\farcs08) cluster C24 (Figure 1).

Thus, the high angular HST resolution allows us to conclude that the
region where SN 2000ft exploded is not identified with any of the 1.1$\mu$m-selected (or optical) clusters detected in the F110W HST image (D\'{\i}az-Santos et a
l. 2007).  According to the HST color map (Figure 1), SN 2000ft appears to have exploded in a region near the edge of a strong lane
 of dust surrounding the ring.  This region could trace the location of a dust-obscured star cluster. The vast
  majority of the  1.1$\mu$m-selected clusters are low extincted (A$_V$ $\sim$ 1 mag), intermediate age ($\sim$
   9-20 Myr) with stellar masses in the 0.1 to 1 $\times$ 10$^7 \Msun$ range, while a few are more extincted
    (A$_V\sim$ 3 mag) and younger (D\'{\i}az-Santos et al. 2007). The apparent optical flux of a cluster similar in
     mass to these, but obscured by a screen of dust with an extinction equivalent to about 4 visual magnitudes,
      would be below the detection limit of the present optical HST images.  Only a very massive (i.e.
mass above 10$^7 \Msun$) cluster would have been detected behind 4 magnitudes of visual extinction. Therefore,  the fact that no cluster is identified at the location of SN 2000ft is consistent with a  dust-obscured parent star cluster having a mass of no more than 10$^7 \Msun$.

 \subsection{Optical Detection of Supernovae in (U)LIRGS}

  Although there is only one ULIRG, Arp 220, at a distance of less than 100 Mpc, the number of
 LIRGs  (i.e. 10$^{11}\Lsun < L_{IR} < 10^{12}\Lsun$) like NGC 7469 within this distance is large.
 Of the 629 galaxies
 in the IRAS Revised Bright Galaxy Sample (Sanders et al. 2003), approximately 200 are in the LIRG
 luminosity range, most of them at distances of about 100 Mpc, or less. The expected supernova rate in
 these LIRGs corresponds to 0.24 - 2.4 yr$^{-1}$, if the infrared luminosity is entirely produced
 by starbursts (Mannucci et al. 2003).

 Recent attempts to detect supernovae in (U)LIRGs with the HST/NICMOS near-IR camera have failed
 (Cresci et al. 2007). Based on the infrared luminosity and number (17) of galaxies in the sample, no less
 than 12 supernovae were expected of which no confirmed event was detected.  According to the authors,
 the shortage of supernova detections could be explained by a combination of effects, the most important
 being the existence of strong extinction (A$_V \geq$ 11 mag.) everywhere within the starburst regions, and the
 size of the region where the supernovae are generated . If supernovae were exploding within the
 inner 500 pc, even NICMOS would not be able to detect them. 

 Although the non detection of supernovae using NICMOS is not very encouraging,
 there are several aspects of the detection strategy that can be improved, therefore increasing the chances of
 detecting a larger fraction of the expected supernovae. It is well known than massive starbursts in (U)LIRGs occur
 mostly in the inner regions of these galaxies, so the sample has to be selected such that the best linear resolution
 is achieved. While the sample of (U)LIRGs selected by Cresci and collaborators is at an average distance of 150
  Mpc, monitoring of other galaxies such as the NICMOS volume-limited sample of nearby LIRGs
 (Alonso-Herrero et al. 2006), at an average distance of 60 Mpc, would be more desirable. The expected number of 
 SNe per year in such a sample would be about 12 for a total of 30 LIRGs and an average luminosity (L$_{IR}$)
 of 2 $\times 10^{11}\Lsun$.

 Regarding internal extinction, although (U)LIRGs are dusty galaxies, the dust distribution in the nuclear regions
 is very patchy presenting regions of high and moderate extinctions on scales of hundred parsecs
  (Alonso-Herrero et al. 2006). The nuclear regions of (U)LIRGs are on average very red with typical I$-$H colors 
   in the 1.5 to 2.5 range (Surace et al. 2000) due to a complex combination of stellar populations, and dust
    absorption and emission.
  Even though the extinction in the I-band is about a factor 2.5 times higher than in the H-band (e.g. Rieke
 \& Lebofsky 1985), the advantage of going to the near-IR where the extinction is lower, is 
 compensated in the optical by the fact type II supernovae have blue intrinsic 
 I$-$H colors of 0.5, or less,  at or near maximum light (Pozzo et al. 2006; di Carlo et al. 2002), i.e. 
 about one to two magnitudes bluer than the nuclear regions of (U)LIRGs.
 Therefore, even if as in NGC 7469, only 10\% of the recent star formation in LIRGs were in
  regions of low to moderate visual extinctions
 (i.e. A$_V \leq$ 5 mag), one could expect to detect several supernova events a year by monitoring a
  representative sample of low redshift LIRGs with HST.
 In this respect, near-IR NIC2 imaging does not provide the best
 angular resolution. Selecting shorter (optical) wavelengths increases the angular
 resolution in diffraction-limited telescopes by a factor of at least two, if for example the I-band is used.
 Moreover, HST optical instruments such as the ACS High Resolution Camera (HRC) has pixels with an angular size
 three smaller than that of NIC2 (0\farcs026 versus 0\farcs0755).  
 The smaller pixelation coupled
 with the shorter wavelengths increases the overall angular resolution of I-band ACS/HRC images by a factor
 of more than two with respect to that of NIC2 H-band images.

 The different size of the pixels implies that the extended galaxy surface brightness seen by ACS/HRC pixels is a
 factor 9 smaller than that seen by NIC2 pixels, and therefore the contrast between the expected supernova
 emission (i.e. high surface brightness point source) and the surrounding, lower surface brightness emission
 will increase accordingly.
 The apparent H-band surface brightness of the inner 1\farcs0 radius region of ULIRGs is in the
 16 $-$ 18 mag arcsec$^{-2}$ range (Colina et al. 2001c). The PSFs of ACS/HRC F814W and NIC2 F160W are
 such that
 80\% of the flux of a point source is within an aperture of 0\farcs15 and 0\farcs3 in radius, respectively. Thus, the
 apparent I- and H-band magnitudes of the galaxy in the corresponding apertures
 are in the 21 $-$ 23.0 and 17.4 $-$ 19.4 ranges, respectively, assuming typical I-H colors of 2 for the nuclear
 regions of ULIRGs (Surace et al. 2000).

 As shown in previous sections the combination of multi-epoch observations with high
 angular resolution has allowed us to detect supernova SN 2000ft in the optical through an
 extinction of at least three magnitudes in the visual. SN 2000ft is classified as a bright radio
 supernova similar in luminosity  to other radio supernovae recently detected in the nucleus of
 Arp 220 (Lonsdale et al. 2006), and falling in the upper end of the range of observed radio peak
 luminosities for known type II supernovae (Alberdi et al. 2006). These
 intrinsically bright supernovae will have apparent red magnitudes of
 m$_I$ $\leq$ 16.5, for distances of less than 100 Mpc, if not affected by extinction. More regular type II 
 supernovae are about
 two magnitudes fainter than their bright counterparts (Patat et al. 1994). Thus, both bright and
 regular type II supernovae would have I- and H-band apparent
 magnitudes 19.5 $\leq$ m$_I \leq$21.5 and 17.5 $\leq$ m$_H \leq$ 19.5, respectively,  if located in regions
 of low to moderate extinction (i.e. A$_V \leq$ 5 mag). Therefore, supernovae in these regions
 could be detected close to maximum light with relatively short (i.e. snapshots) I-band ACS exposures,
 and more easily than in the near-IR H-band as the expected galaxy flux contribution is of the order of that
 of the supernova. Future HST instruments like the WFPC3 UVIS channel with pixels of 0\farcs040 (i.e. 1.6 
 times ACS/HRC) and a throughput higher than that of ACS/HRC in the optical range are also adecuate for
 the detection of supernovae as explained above. 

  In summary, multi-epoch, high angular ($\sim$ 0$\farcs$1) resolution optical (red) imaging
  should be considered potentially as a valid technique to probe the presence of supernovae in
  moderately absorbed star-forming regions in (U)LIRGs. Recently, adaptive optics (AO) near-infrared 
  imaging (angular resolution of 0\farcs08 with pixels of 0\farcs027) with the ESO Very Large Telescope 
  AO system,  has allowed to detect a supernova (SN 2004ip) in the nuclear region of  the LIRG 
  IRAS 18293$-$3413, at a distance of 500 pc from its nucleus (Mattila et al. 2007), i.e. about the same 
  distance from the nucleus as SN 2000ft in NGC 7469 (Mattila et al. 2007). 
   Detection and characterization of supernovae in the
  highly absorbed regions (A$_V >>$ 5 mag) will be accessible through mid-IR, or radio multi-epoch
  imaging as already demonstrated (Alberdi et al. 2006; Lonsdale et al. 2006, Beswick 2007 for a review).

\section{Conclusions}

     Multi-epoch Hubble Space Telescope (HST) imaging has been used to search for the 
     optical emission of the luminous radio supernova SN 2000ft, previously detected at 
     radio frequencies in the circumnuclear star-forming ring of the luminous infrared, 
     Seyfert 1 galaxy NGC 7469. The main conclusions of this study are:
     
  \begin{enumerate}
     
  \item SN 2000ft has been detected in the optical 
     in two independent Planetary camera images taken on May 13, 2000,  about two months 
     ($\pm$ 40 days) before the predicted explosion date based on the analysis of its multi-year radio light 
     evolution.  
     
   \item The extinction corrected absolute optical magnitudes place SN 2000ft in the range of
     luminous optical type II supernovae, as other luminous radio supernovae such as SN 1979C 
     and SN 1988Z before.  
     
    \item SN 2000ft exploded in a region located at only 0\farcs1 (i.e. 34 $\pm$ 3 pc) west of a 
      faint cluster (C24). No parent cluster is identified in the HST images.
      
     \item The unambiguous detection of SN 2000ft behind at least three magnitudes of extinction in
      the visual shows that multi-epoch sub-arcsecond (FWHM $\sim$ 0\farcs1) optical imaging is also a valid tool
     to detect supernovae even in the dusty (circum)nuclear regions of (ultra)luminous infrared galaxies.

   \end{enumerate}

\begin{acknowledgements}
      AAH, LC and TDS acknowledge support by the Spanish Plan Nacional del
     Espacio under grant ESP2005-01480. AA was supported by grant
     AYA2005-08561-C03-02. JMT was partially supported by grant AYA2005-08523-C03.
     Discussions and support with members of the WFPC2 team, Shireen Gonzaga and Ray Lucas
     are acknowledged.
\end{acknowledgements}

\end{document}